\documentclass{elsart}

\usepackage{graphicx}
\usepackage{epsfig,array}
\usepackage{booktabs}

\newcommand{\beq}{\begin{equation}}
\newcommand{\eeq}{\end{equation}}
\newcommand{\beqa}{\begin{eqnarray}}
\newcommand{\eeqa}{\end{eqnarray}}

\newcommand{\sla}[1]%
        {\kern .25em\raise.18ex\hbox{$/$}\kern-.75em #1}
\newcommand{\mybar}[1]%
        {\kern 0.8pt\overline{\kern -0.8pt#1\kern -0.8pt}\kern 0.8pt}

\begin{document} 
\begin{frontmatter}
\title{Higgs-Mediated $\tau\rightarrow \mu$ and $\tau\rightarrow e$ transitions
in II Higgs doublet Model and Supersymmetry}
\author[romeII]{P.Paradisi}
\address[romeII]{University of Rome ``Tor Vergata'' and INFN sez. RomaII, 
	Via della Ricerca Scientifica 1, 
	I-00133 Rome}

\begin{abstract}
We study the phenomenology of the $\mu-\tau$ and $e-\tau$ lepton flavour
violation (LFV) in a general two Higgs Doublet Model (2HDM) including the
supersymmetric case.
We consider several LFV decay modes of the charged fermion $\tau$, namely 
$\tau\rightarrow l_j\gamma$, $\tau\rightarrow l_j l_k l_k$ and 
$\tau\rightarrow l_j\eta$. The predictions and the correlations among 
the rates of the above processes are computed.
In particular, it is shown that $\tau\rightarrow l_j\gamma$ processes are 
the most sensitive channels to Higgs-mediated LFV specially
if the splitting among the neutral Higgs bosons masses is not below the 
$10\%$ level.
 
\end{abstract}

\end{frontmatter}

\section{Introduction}
\label{sec:introduction}

The observation of neutrino oscillation have established the existence of 
lepton family number violation.
As a natural consequence of this phenomenon, one would expect flavour mixing to appear also in the charged leptons sector.
This mixing can be manifested in rare decay processes such as 
$\mu\rightarrow e\gamma$, $\tau\rightarrow \mu\gamma$ etc. 
In the Standard Model with massive neutrinos these processes are mediated, at one loop level, by the exchange of the $W$ bosons and neutrinos; however, in analogy to the quark
sector, the resulting rates are GIM suppressed and turn out to be proportional to the ratio of masses of neutrinos over the masses of the W bosons. In addition, if neutrinos are
massive, we would expect LFV transitions also in the Higgs sector
through the decay modes $H^0\rightarrow l_i l_j$ mediated at one loop level by the exchange of the $W$ bosons and neutrinos.
However, as for the $\mu\rightarrow e\gamma$ and the $\tau\rightarrow \mu\gamma$ case, also the $H^0\rightarrow l_i l_j$ rates are GIM suppressed.\\In a supersymmetric (SUSY) framework the situation is completely dif\-fe\-rent.
Besides the previous contributions, supersymmetry provides new direct sources of flavour violation, namely the possible presence of off-diagonal soft terms in the slepton mass matrices and in the trilinear couplings \cite{fbam}. In practice, flavour violation would 
originate from any misalignment between fermion and sfermion mass eigenstates.
LFV processes arise at one loop level through the exchange of neutralinos (charginos) and charged sleptons (sneutrinos). The amount of the LFV is regulated by a Super-GIM mechanism that can be much less severe than in the non supersymmetric case 
\cite{Masierorew,textures}.
\footnote{As recently shown in ref.\cite{mfvl}, some of these effects are common to many extensions of the SM, even to non-susy scenarios, and can be described in a general way in terms of an effective field theory.}\\
Another potential source of LFV in models such as the minimal
supersymmetric standard model (MSSM) could be the Higgs sector, in fact, 
extensions of the Standard Model containing more than one Higgs 
doublet generally allow flavor-violating couplings of the neutral Higgs bosons.
Such couplings, if unsuppressed, will lead to large flavor-changing neutral currents in direct opposition to experiments. The MSSM avoid these dangerous couplings at the tree level segregating the quark and Higgs fields so that one Higgs $(H_u)$ can couple only to up-type quarks while the other $(H_d)$ couples only to d-type.
Within unbroken supersymmetry this division is completely natural,
in fact, it is required by the holomorphy of the superpotential.
However, after supersymmetry is broken, couplings of the form $QU_cH_d$
and $QD_cH_u$ are generated at one loop \cite{Hrs}.
In particular, the presence of a non zero $\mu$ term, coupled with SUSY breaking, is enough to induce non-holomorphic Yukawa interactions for quarks and leptons. 
For large $\tan\beta$ values the contributions to d-quark masses coming from non-holomorphic operator  $QD_cH_u$ can be equal in size to those coming from the usual holomorphic operator $QD_cH_d$ despite the loop suppression suffered by the former. 
This is because the operator itself gets an additional enhancement of $\tan\beta$.\\ 
As shown in reference \cite{bkq} the presence of these loop-induced non-holomorphic couplings also leads to the appearance of flavor-changing couplings of the neutral Higgs bosons. These new couplings generate a variety of flavor-changing processes such as 
$B^0 \rightarrow\mu^{+}\mu^{-}$, $\bar{B^0}-B^0$ etc.\cite{isiburapila}.\\
Higgs-mediated FCNC can have sizable effects also in the lepton sector
\cite{bkl}: given a source of non-holomorphic couplings, and LFV among the sleptons,
Higgs-mediated LFV is unavoidable.
These effects have been widely discussed in the recent literature
both in a generic 2HDM \cite{chang1,chang} and in supersymmetry \cite{sher} frameworks. However, so far most of the attention has been devoted to the tree level effects and in particular to the $\tau\rightarrow l_jl_kl_k$ and $\tau\rightarrow \mu\eta$ processes.
On the other hand, the Higgs-mediated FCNC can have a sizable impact also in loop-induced processes, such as $\tau\rightarrow l_j\gamma$.
The main purpose of this letter is a detailed investigation of these effects
(a comprehensive analysis of the $e-\mu$ transitions 
will be presented in an upcoming letter \cite{hlfvemu}).
We consider, in particular, the additional dipole and monopole operators 
induced by the Higgs exchange. 
As a consequence, $\tau\rightarrow l_j\gamma$ processes are ge\-ne\-ra\-ted and 
$\tau\rightarrow l_jl_kl_k$ decay rates get additional contributions by the monopole 
and dipole operators.
We perform the analysis both in a general and in a supersymmetric two Higgs Doublet Models.

\section{LFV in the Higgs Sector}

As it is well known, Standard Model extensions containing more than one Higgs 
doublet generally allow flavor-violating couplings of the neutral Higgs bosons
which arise as a consequence of the fact that each fermion type can couple to both Higgs doublets. Such couplings, if unsuppressed, will lead to large flavor-changing neutral currents in direct opposition to experiments.
The possible solution to this problem involve an assumption about the Yukawa structure of the model. A discrete symmetry can be invoked to allow a given fermion type to couple to a single Higgs doublet, and in such case FCNC's are absent at tree level. In particular, when a single Higgs field gives masses to both types of fermions the resulting model is referred as 2HDM-I. On the other hand, when each type of fermion couples to a different Higgs doublet the model is said 2HDM-II.\\
When each fermion type couple to both Higgs doublets, FCNC could be kept under control if there exists a hierarchy among the Yukawa matrices. For instance, it is possible to assume that the model has a flavor symmetry able to reproduce the observed fermion masses and mixing angles.
Another possibility is that each type of fermion couples to a different Higgs doublet at the tree level, and the coupling with the other Higgs doublet arises only as a radiative effect. In the following we will assume the last scenario.
This occurs, for instance, in the MSSM where the type-II 2HDM structure 
is not protected by any symmetry and is broken by loop effects.\\
We consider the following generic Yukawa interactions for charged leptons, including the radiatively induced LFV terms:
\beq
\mathcal{-L}\simeq
Y_{\mu}H^{0}_{1}\overline{\mu}_R\mu_L-Y_{\tau}H^{0}_{1}\overline{\tau}_R\tau_L
\!+\!Y_{\tau}H^{0}_{2}\Delta^{3j}_{L}\overline{\tau}_R l^j_L
\!+\!Y_{\tau}H^{0}_{2}\Delta^{3j}_{R}\overline{l^j_R}\tau_L + h.c.
\eeq
where the $\Delta^{3j}_{L,R}$ parameters are the source of LFV
(for instance, in the MSSM, they are generated at one loop level by the slepton mixing).\\
In the mass-eigenstate basis for both leptons and Higgs bosons,
the effective flavor-violating interactions are described by the four dimension operators:
\beqa
\mathcal{-L}&\simeq&(2G_{F}^2)^{\frac{1}{4}}\frac{m_{l_i}} {c^2_{\beta}}
\left(\Delta^{ij}_{L}\overline{l}^i_R l^j_L+\Delta^{ij}_{R}\overline{l}^i_L l^j_R \right)
\left(c_{\beta-\alpha}h^0-s_{\beta-\alpha}H^0-iA^0 \right)\nonumber\\\nonumber\\
&+&
(8G_{F}^2)^{\frac{1}{4}}\frac{m_{l_i}} {c^2_{\beta}}
\left(\Delta^{ij}_{L}\overline{l}^i_R \nu^j_L+\Delta^{ij}_{R}\nu^i_L\overline{l}^j_R\right)
H^{\pm} + h.c.\nonumber
\eeqa
where $\alpha$ is the mixing angle between the CP-even Higgs bosons $h_0$ and $H_0$, $A_0$ is the physical CP-odd boson, $H^{\pm}$ are the physical charged Higgs-bosons and $\tan\beta$ is the ratio of the vacuum expectation value for the two Higgs. Irrespective to the mechanism of the high energy theories generating the LFV, we treat the $\Delta^{ij}_{L,R}$ terms in a model independent way
\footnote{On the other hand, there are several models with a specific ansatz about the flavour-changing couplings. For instance, the famous multi-Higgs-doublet models proposed by Cheng and Sher \cite{chengsher} predict that the LFV couplings of all the neutral Higgs bosons with the fermions have the form $Hf_if_j \sim \sqrt{m_im_j}$.}.
In order to constrain the $\Delta^{ij}_{L,R}$ parameters, we impose that their contributions to LFV processes as $l_i\rightarrow l_jl_kl_k$ and $l_i\rightarrow l_j\gamma$ do not exceed the experimental bounds.
At tree level, Higgs exchange contribute only to $l_i\rightarrow l_jl_kl_k$. 
On the other hand, at the one loop level, also the dipole operators arise 
and the LFV radiative decays $l_i\rightarrow l_j\gamma$ are allowed.
However, the one loop higgs mediated dipole transition implies three chirality
flips: two in the Yukawa vertices and one in the lepton propagator.
This strong suppression can be overcome at higher order level. 
Going to two loop level, one has to pay the typical price of $g^2/16\pi^2$ 
but one can replace the light fermion masses from yukawa vertices with 
the heavy fermion (boson) masses circulating in the second loop.
In this case, the virtual higgs boson couple only once to the lepton
line, inducing the needed chirality flip.
As a result, the two loop amplitude can provide the major effects.
Naively, the ratio between the two loop fermionic amplitude and the one loop amplitude is:
$$
\frac{A^{(2-loop)_{f}}_{l_i\rightarrow l_j\gamma}}{A^{1-loop}_{l_i\rightarrow l_j\gamma}}
\sim
\frac{\alpha_{em}}{4\pi}\frac{m^2_{f}}{m^2_{l_i}}\log\bigg(\frac{m^2_{f}}{m^2_{H}}\bigg)
$$ 
where $m_{f}= m_{b}, m_{\tau}$ is the mass of the heavy fermion circulating in the loop. 
We remind that in a Model II 2HDM the Yukawa couplings between 
neutral Higgs bosons and quarks are $H\bar{t} t \sim m_t/\tan\beta$ and 
$H\bar{b} b \sim m_b \rm{\tan\beta}$.
Since the Higgs mediated LFV is relevant only at large $\tan\beta\geq 30$,
it is clear that the main contributions arise from the $\tau$ and 
$b$ fermions and not from the top quark.
So, in this framework, $\tau\rightarrow l_j\gamma$ do not receives sizable
two loop effects by an heavy fermionic loop differently from the
$\mu\rightarrow e\gamma$ case.
However, the situation can drastically change when a $W$ boson circulates
in the two loop Barr-Zee diagrams.
Bearing in mind that $H W^{+}W^{-}\sim m_W$ and that 
pseudoscalar bosons do not couple to a $W$ pair, it turns out that
$A^{(2-loop)_{W}}_{l_i\rightarrow l_j\gamma}/A^{(2-loop)_{f}}_{l_i\rightarrow l_j\gamma}\sim m^2_{W}/(m^2_{f} tan\beta )$ thus, two loop $W$ effects
are expected to dominate, as it is confirmed numerically \cite{chang1}.\\
Moreover, up to one loop level, $l_i\rightarrow l_jl_kl_k$ gets 
additional contributions induced by $l_i\rightarrow l_j\gamma^*$ amplitudes.
It is worth noting that the Higgs mediated monopole and dipole amplitudes have 
the same $\tan^3\beta$ dependence. This has to be contrasted to the non-Higgs contributions. For instance, within susy, the gaugino mediated dipole amplitude
is proportional to $\tan\beta$ while the monopole amplitude is $\tan\beta$ independent.\\
The general expression for the Higgs mediated $l_i\rightarrow l_jl_kl_k$ and $l_i\rightarrow l_j\gamma$ rates read:
\beqa
\frac{Br(l_i\rightarrow l_jl_kl_k)}{Br(l_i\rightarrow l_j\nu_i\nu_j)}&=& \frac{1}{8 G_F^2}
\bigg[(3+5\delta_{jk})\frac{|S^{+}_{L}|^2}{8}+(3+\delta_{jk})\frac{|S^{-}_{L}|^2}{4}+
e^4\left|M_{L}\right|^2(2+\delta_{jk})\nonumber\\\nonumber\\
&-&4e^4D^{\gamma}_LM_{L}(2+\delta_{jk})
+ 8e^4\left|D^{\gamma}_L\right|^2\left(\log\frac{m^2_{l_i}}{m^2_{l_k}}-3\right)+
(L\leftrightarrow R)\bigg]\nonumber\, 
\eeqa
$$
\frac{Br(l_i\rightarrow l_j\gamma)}{Br(l_i\rightarrow l_j\nu_i\nu_j)}=
\frac{48 \pi^3\alpha_{em}}{G_F^2}
\bigg[\left|D^{\gamma}_L\right|^2+\left|D^{\gamma}_R\right|^2\bigg]
$$
where the scalar $S_{L,R}$, the monopole $M_{L,R}$ and the dipole $D_{L,R}$ 
amplitudes read: 
\beq
S_{L,R}^{+} = \frac{G_F}{\sqrt2}\frac{m_{l_i}m_{l_k}}{M^2_H}
\frac{1}{c^3_{\beta}}\bigg[\frac{c_{\alpha}s_{\beta-\alpha}}{m^2_H}
-\frac{s_{\alpha}c_{\beta-\alpha}}{m^2_h}+\frac{s_{\beta}}{m^2_A}\bigg]
\,\Delta_{L,R}\,\, 
\eeq
\beq
S_{L,R}^{-} = \frac{G_F}{2}\frac{m_{l_i}m_{l_k}}{M^2_H}
\frac{1}{c^3_{\beta}}\bigg[\frac{c_{\alpha}s_{\beta-\alpha}}{m^2_H}
-\frac{s_{\alpha}c_{\beta-\alpha}}{m^2_h}-\frac{s_{\beta}}{m^2_A}\bigg]
\,\Delta_{L,R}\,\,  
\eeq
\beqa
M_{L,R} = \frac{G_F}{48\sqrt2\pi^2}\frac{m^2_{l_i}}{c^3_{\beta}}
&\bigg[&\frac{c_{\alpha}s_{\beta-\alpha}}{m^2_H}\left(\log\frac{m^2_{l_i}}{M^2_H}+\frac{5}{6}\right)-
\frac{s_{\alpha}c_{\beta-\alpha}}{m^2_h}\left(\log\frac{m^2_{l_i}}{M^2_h}+\frac{5}{6}\right)
\nonumber\\\nonumber\\
&+&
\frac{s_{\beta}}{m^2_A}\left(\log\frac{m^2_{l_i}}{M^2_A}+\frac{5}{6}\right)\bigg]\Delta_{L,R}
\eeqa
\beqa
D_{L} &=& -\frac{G_F}{8\sqrt2\pi^2}\frac{m^2_{l_i}}{c^3_{\beta}}
\bigg[\frac{c_{\alpha}s_{\beta-\alpha}}{m^2_H}\left(\log\frac{m^2_{l_i}}{M^2_H}+
\frac{4}{3}-\frac{\alpha_{el}}{\pi}\frac{m^{2}_W}{m^{2}_{\tau}}
\frac{F(a_{W})}{\tan\beta}\right)+\nonumber\\\nonumber\\
&-&
\frac{s_{\alpha}c_{\beta-\alpha}}{m^2_h}\left(\!\log\frac{m^2_{l_i}}{M^2_h}\!+\!
\frac{4}{3}\!-\!\frac{\alpha_{el}}{\pi}
\frac{m^{2}_W}{m^{2}_{\tau}}
\!\frac{F(a_{W})}{\tan\beta}\right)\!-\!
\frac{s_{\beta}}{m^2_A}\left(\log\frac{m^2_{l_i}}{M^2_A}\!+\!
\frac{5}{3}\right)\bigg]\Delta_{L}
\eeqa
\beq
D_{R} = D_{L}(L\leftrightarrow R) +\frac{G_F}{48\sqrt2\pi^2}
\frac{m^2_{l_i}}{M^2_{H^{\pm}}}\frac{\Delta_{R}}{c^3_{\beta}}\,.
\eeq
where $a_{W}=m^{2}_{W}/m^{2}_{H}$. The terms proportional to $F(a_W)$ arise 
from two loop effects induced by Barr-Zee type diagrams with a $W$ boson exchange.
The loop function $F(z)$ is given by
\beq
F(z)\simeq 3f(z)+\frac{23}{4}g(z)+\frac{f(z)-g(z)}{2z}
\eeq 
with the Barr-Zee loop integrals given by:
\beq
g(z) = \frac{1}{4}\int_0^1 dx \frac{\log{(z/x(1-x))}}{z-x(1-x)}\,,
\eeq
\beq
f(z) = \frac{1}{4}\int_0^1 dx \frac{1-2x(1-x)\log{(z/x(1-x))}}{z-x(1-x)}.
\eeq
For $z\ll 1$ it turns out that:
\beq
F(z)\sim \frac{35}{16}(\log z)^{2}+\frac{\log z+2}{4z}.
\eeq
The $\tau\rightarrow \mu(e)\eta$ process receives the only contribution from the pseudoscalar A and the resulting branching ratio is:
$$
\frac{Br(\tau\rightarrow l_j\eta)}{Br(\tau\rightarrow l_j\bar{\nu_j}\nu_{\tau})} \simeq
9\pi^2\!\left(\frac{f^{8}_{\eta} m^{2}_{\eta}}{m^{2}_{A}m_{\tau}}\right)^{\!2}\!
\!\left(1\!-\!\frac{m^{2}_{\eta}}{m^{2}_{\tau}}\right)^{\!2}\!
\!\left[\xi_s\!+\!\frac{\xi_b}{3}\!\left(1\!+
\!\sqrt2\frac{f^{0}_{\eta}}{f^{8}_{\eta}}\right)\right]^2
\!\!\!\Delta_{3j}^2\tan^6\beta
$$
where $m^{2}_{\eta}/m^{2}_{\tau}\simeq 9.5\times10^{-2}$ and the relevant 
decay constants are $f^{0}_{\eta}\sim 0.2 f_{\pi}$,
$f^{8}_{\eta}\sim 1.2 f_{\pi}$ and $f_{\pi}\sim 92$ MeV \cite{fedelman}.
The parameters $\xi_{f}$ appear in the couplings between the scalar and the fermions
$-i(\sqrt 2 G_F)^{1/2}\tan\beta H\xi_{f} m_f\overline{f}f$.
Although they are equal to one at tree level
they can get large corrections from higher order effects.
This is the case, for instance, of Susy where contributions 
arising from gluino-squark loops (proportional to $\alpha_{s}\!\tan\beta$)
can enhance or suppress significantly the tree level value of $\xi_{b}$
\cite{Hrs,bkq,isiburapila}.\\

\subsection{Non-decoupling limit: $\sin(\beta-\alpha)=0$}

In this section we will derive the expressions and the correlations
among the rates of the above 
processes in the limiting case where $\sin(\beta-\alpha)=0$ and
$\tan\beta$ is large. In particular, we will establish which the most
promising channels to detect Higgs mediated LFV are.\\
For $\tau\rightarrow l_j\gamma$ and $\tau\rightarrow l_j l_k l_k$
branching ratios we get, respectively

\beq
\frac{Br(\tau\rightarrow l_j\gamma)}{Br(\tau\rightarrow l_j\bar{\nu_j}\nu_\tau)}\simeq
\frac{3\alpha_{el}}{8\pi}\left(\frac{m^2_{\tau}}{m^2_{h,A}}\right)^2
\left(\log{\frac{m^2_{\tau}}{m^2_{h,A}}}+\frac{4}{3}\right)^2\tan^6\beta
\Delta_{3j}^2
\eeq

\beqa
\frac{Br(\tau\rightarrow l_jl_kl_k)}{Br(\tau\rightarrow l_j\bar{\nu_j}\nu_\tau)} &\simeq&
\frac{m^2_{\tau}m^2_{l_k}}{8m^4_{h,A}}
\Delta_{3j}^2\tan^6\beta
\bigg[\frac{3}{8}\left(1+\delta_{jk}\right)+
\frac{\alpha^{2}_{el}}{\pi^2}\frac{m^2_{\tau}}{m^2_{l_k}}\nonumber\\\nonumber\\
&\cdot&
\left(\log\frac{m^2_{\tau}}{m^2_{l_k}}\!-\!3\right)\left(\log\frac{m^2_{\tau}}{m^2_{h,A}}+\frac{4}{3}\right)^{\!2}\bigg]
\eeqa

where we have retained only the dominant contribution from the lightest $h$ or $A$ Higgs bosons. In the above expressions we disregarded subleading two loop effects although they are retained in the numerical analysis.
On the other hand, two loop effects provide a sizable reduction of 
$Br(\tau\rightarrow l_j\gamma)$ and $Br(\tau\rightarrow l_jee)$
in the large $m_h$ regime as it is shown in fig. 1.
Such effects are not visible in $Br(\tau\rightarrow l_j\mu\mu)$
because it is dominated by the tree level Higgs exchange contributions.
We note that, while $\tau\rightarrow\mu(e)\eta$ rates decouple
in the heavy pseudoscalar limit, the $Br(\tau\rightarrow l_jl_kl_k)$ and
$Br(\tau\rightarrow l_j\gamma)$
branching ratios can get additional contributions by the $h$ scalar.
The $\tau\rightarrow l_jl_kl_k$ rates contain two terms: the first comes from the tree level Higgs exchange, the second from the dipole operator
neglecting subdominant contributions by the monopole operator.\\
In general the one loop induced Higgs contributions have both advantages and disadvantages. The disadvantages consist in the additional $\alpha^2_{el}$ factor, the advantages consist in the possibility to replace light lepton masses with the mass of the decaying particles. In addition, we get an extra large $\log(m^2_{\tau}/m^2_{h})$ 
factor from the loop functions. We remark that the scalar contributions
to $\tau\rightarrow l_j ee$ are very suppressed compared to the dipole contributions while they are of the same order in the $\tau\rightarrow l_j\mu\mu$ cases.\\
In order to understand which the best candidate to detect LFV among 
$\tau\rightarrow l_jl_kl_k$, $\tau\rightarrow l_j\gamma$ or $\tau\rightarrow l_j \eta$ is, we derive the following relations:
\beq
\frac{Br(\tau\!\rightarrow\!l_j\gamma)}{Br(\tau\!\rightarrow\!l_j\eta)}\simeq\, 
\frac{1}{5}\left(\log\frac{m^2_{\tau}}{m^2_{A}}+\frac{4}{3}\right)^2 \sim 10
\eeq
\beq
\frac{Br(\tau\!\rightarrow\!l_j ee)}{Br(\tau\!\rightarrow\!l_j\gamma)} \simeq\, 
\frac{\alpha_{el}}{3\pi}\left(\!\log\frac{m^2_{\tau}}{m^2_{e}}\!-\!3\!\right)\sim 10^{-2}
\eeq
\beqa
\frac{Br(\tau\!\rightarrow\!l_j\mu\mu)}{Br(\tau\!\rightarrow\!l_j\gamma)}
&\simeq\,& 
\frac{\alpha_{el}}{3\pi}\left(\!\log\frac{m^2_{\tau}}{m^2_{\mu}}\!-\!3\!\right)
+
\frac{\pi}{\alpha_{el}}\,\frac{(1+\delta_{j\mu})}{8} 
\bigg(\frac{m^2_{\mu}}{m^2_{\tau}}\bigg)
\bigg(\!\log\frac{m^2_{\tau}}{m^2_{A}}\!+\!\frac{4}{3}\bigg)^{\!-2}
\nonumber\\\nonumber\\
&\sim&
\,\bigg[2+3(1+\delta_{j\mu})\bigg]\cdot 10^{-3}
\eeqa
where the last equalities in the above equations are obtained by setting 
$m_A=150 \rm{GeV}$. In general, the above equations imply that 
$\tau\rightarrow l_j \gamma$ is dominant with respect to $\tau\rightarrow l_jl_kl_k$ 
or $\tau\rightarrow l_j \eta$ in the not decoupling limit.
In addition, we stress that a tree level Higgs exchange predicts that
$Br(\tau\rightarrow l_j\mu\mu)/Br(\tau\rightarrow l_j ee)
\sim m^2_{\mu}/m^2_e$ while at the one loop level one gets:
\beqa
\frac{Br(\tau\!\rightarrow\!l_j \mu\mu)}{Br(\tau\!\rightarrow\!l_j ee)}
&\simeq\,&
0.2 + 15 (1+\delta_{j\mu})
\bigg(\!\log\frac{m^2_{\tau}}{m^2_{A}}\!+\!\frac{4}{3}\bigg)^{\!-2}
\nonumber\\\nonumber\\
&\sim& 
\bigg[2+3(1+\delta_{j\mu})\bigg]\cdot 10^{-1}.
\eeqa
where the last relation in eq.16 holds for $m_A=150 \rm{GeV}$.
In particular, eq.16 allow us to conclude that, in the not decoupling limit,
$\tau\rightarrow l_j ee$ is more sensitive to Higgs mediated LFV than 
$\tau\rightarrow l_j\mu\mu$, as it is reproduced by fig.1.

\begin{figure}[center]
\begin{tabular}{cc}
\includegraphics[scale=0.35]{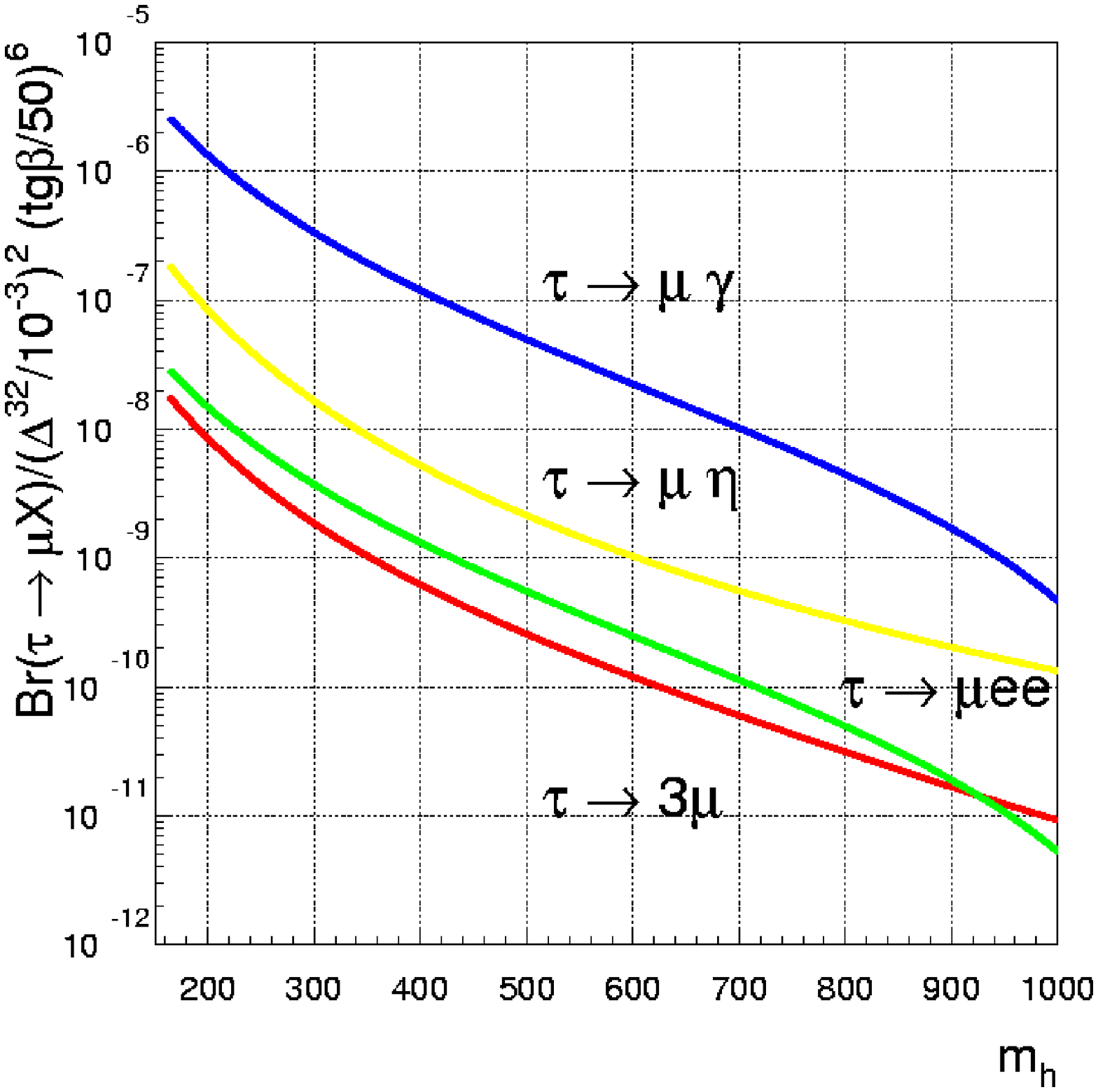} &
\includegraphics[scale=0.35]{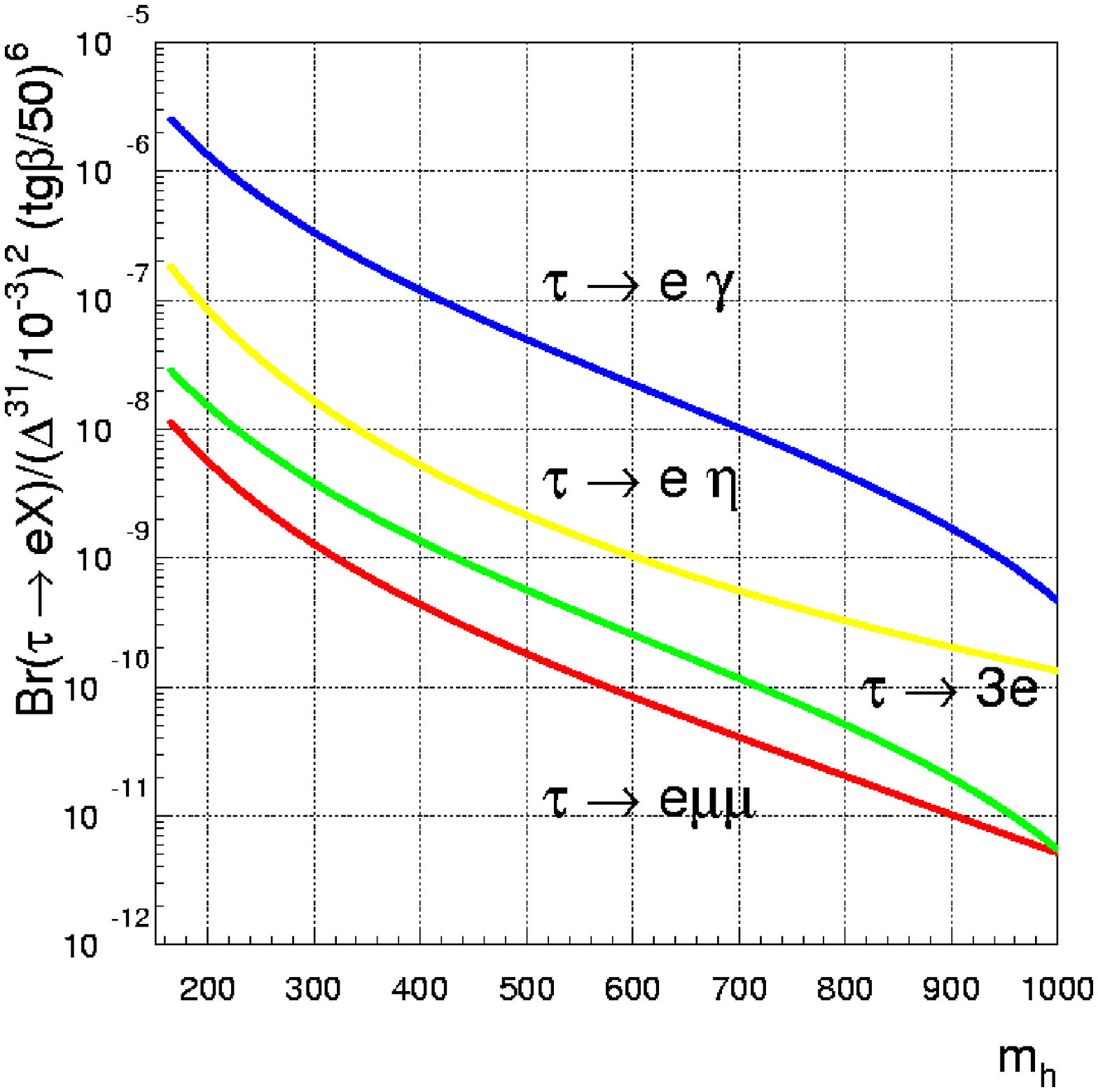}
\end{tabular}
\caption{Branching ratios of various $\tau\rightarrow\mu$ and
$\tau\rightarrow e$ LFV processes vs the lightest Higgs boson
mass $m_h$ in the non decoupling limit.
In the figures we assume $X= \gamma, \mu\mu, ee, \eta$.}
\end{figure}


\subsection{Decoupling limit: $\cos(\beta-\alpha)=0$}

In the decoupling limit, where $\cos(\beta-\alpha)=0$ and 
$m_Z/m_{A^0}\rightarrow 0$, the couplings of the light Higgs boson $h^0$ 
are nearly equal to those of the SM Higgs boson.
This is a particularly interesting limit being that achieved in the Susy framework. In the decoupling limit, $m_{A^0}\simeq m_{H^0}\simeq m_{H^{\pm}}$ (the mass differences are of order $\mathcal{O}(m^2_Z/m_{A^0})$) and, in particular, the MSSM predicts \cite{haber}:
\beq
m^2_{A}-m^2_{H}= \frac{\alpha_2N_c\mu^2}{24\pi M^4_{susy}}
\left(\frac{A^2_t m^4_t}{s^4_{\beta}m^2_W}+
\frac{A^2_b m^4_b}{c^4_{\beta}m^2_W}\right)
\eeq
where $A_{t,b}$ are parameters appearing in the trilinear scalar couplings, $\mu$ is the mixing mass between the two Higgs in the superpotential and $M_{susy}$ is a typical susy scalar mass.
It turns out that pseudoscalar and scalar one loop amplitudes
have opposite signs so, being $m_A\simeq m_H$, they cancel each other 
to a very large extent.
Since these cancellations occur, two loop effects can become important or even dominant in contrast to the non-decoupling limit case.
As final result, we find the following approximate expressions:
\beqa\underline{}
\frac{Br(\tau\rightarrow l_j\gamma)}{Br(\tau\rightarrow l_j\bar{\nu_j}\nu_{\tau})}
\simeq
\frac{3}{2}\frac{\alpha_{el}}{\pi}\left(\frac{m^2_{\tau}}{m^2_A}\right)^2
\tan^{\!6}\!\beta\Delta^2_{\tau j}
&\bigg(&
\frac{\delta m}{m_{A}}\log{\frac{m^2_{\tau}}{m^{2}_{A}}}+\frac{1}{6}+
\nonumber\\\nonumber\\
&+&\frac{\alpha_{el}}{\pi}\,\bigg(\frac{m^{2}_W}{m^{2}_{\tau}}\bigg)\,
\frac{F(a_{W})}{\tan\beta}
\bigg)^2
\eeqa

\beqa
\frac{Br(\tau\rightarrow l_jl_kl_k)}{Br(\tau\rightarrow l_j\bar{\nu_j}\nu_{\tau})} 
&\simeq&
\frac{m^2_{\tau}m^2_{l_k}}{32m^4_{A}}\Delta_{\tau j}^2\tan^6\beta
\bigg[3+5\delta_{jk}\bigg]+\nonumber\\\nonumber\\
&+&\frac{\alpha_{el}}{3\pi}
\bigg(\log\frac{m^2_{\tau}}{m^2_{l_k}}\!-\!3\bigg)
\frac{Br(\tau\rightarrow l_j\gamma)}{Br(\tau\rightarrow l_j\bar{\nu_j}\nu_{\tau})}
\eeqa
where $\delta m=m_{H}-m_{A}$.

It is noteworthy that one and two loop amplitudes have the same signs.
In addition, two loops effects dominate in large portions of the parameter space, specially for large $m_H$ values, where the mass splitting 
$\delta m=m_{H}-m_{A}$ decreases to zero.
In fig.2 we scan over the $\delta m/m_{A}$ range allowed by the
$A_t,\, A_b,\, \mu,\, M_{susy}$ parameters within
$A_t\!\!=\!A_b\!\!=\!0$ (degenerate case) and $A_t\!=\!A_b\!=\!\mu\!=\!2M_{susy}$.
This choice of the parameter space is phe\-no\-me\-no\-lo\-gi\-cal\-ly 
available and, in particular, it is compatible with the experimental bounds
on the lightest stop and Higgs boson masses.\\ 
To get a feeling of the allowed rates for Higgs-mediated
LFV decays in Supersymmetry it is useful to specify the
$\Delta_{3j}$ expressions in terms of the susy parameters.
We remind that the $\Delta_{3j}$ terms are induced at one loop level by 
the exchange of gauginos and sleptons.
Assuming that all the susy particles are of the same order of magnitude but $\mu$ 
($\mu$ being the Higgs mixing parameter), it turns out that
$$
\Delta_{3j}\sim \frac{\alpha_{2}}{24\pi} \frac{\mu}{m_{SUSY}}\,\, \delta_{3j},
$$
where $\delta_{3j}$ is the LFV insertion in the slepton mass matrices.
The above expression depends only on the ratio of the susy mass scales and it does not decouple for large $m_{SUSY}$.\\ 
The unknown $\delta_{3j}$ parameters can be determined only if we specify completely the LFV susy model.
In fig. 2 we have taken into account the normalization $\Delta_{3j}= 10^{-3}$ that requires, in general, large $\delta_{3j}\sim 1$.
The amount of the $\delta_{3j}$ mass insertions is constrained 
by the gaugino mediated LFV and, in general, $\delta_{3j}\sim 1$ requires $m_{SUSY}\sim$ 1TeV  not to exceed the experimental bounds \cite{1mio}.\\ 
The numerical results shown in fig. 2 allow us to draw several interesting observations:
\begin{itemize}
\item
$\tau\rightarrow l_j\gamma$ has the largest branching ratios except for a region around $m_H\sim 700 \rm{Gev}$ where strong
cancellations among two loop effects sink their size
\footnote{For a detailed discussion about the origin of these cancellations
and their connection with non-decoupling properties of two loop 
$W$ amplitude, see ref.\cite{chang1}.}.
The following approximate relations are found:
$$
\frac{Br(\tau\rightarrow l_j\gamma)}{Br(\tau\rightarrow l_j\eta)} 
\simeq
\left(\frac{\delta m}{m_A}\log\frac{m^2_{\tau}}{m^2_A}\!+\!\frac{1}{6}\!+\!
\frac{\alpha_{el}}{\pi}\,\bigg(\frac{m^{2}_W}{m^{2}_{\tau}}\bigg)\,
\frac{F(a_{W})}{\tan\beta}\right)^2\geq 1\,,
$$
where the last relation is easily obtained by using the approximation for
$F(z)$ given in eq.10. If two loop effects were disregarded, then we would obtain
$Br(\tau\rightarrow l_j\gamma)/Br(\tau\rightarrow l_j\eta)\in(1/36,1)$ 
for $\delta m/m_A \in(0,10\%)$.
Two loop contributions significantly enhance $Br(\tau\rightarrow l_j\gamma)$ specially for $\delta m/m_A\rightarrow 0$.\\
\item
In fig. 2, non negligible mass splitting $\delta m/m_{A}$ effects
can be visible at low $m_H$ regime through the bands of the 
$\tau\rightarrow l_j\gamma$ and $\tau\rightarrow l_jee$ processes.
These effects tend to vanish with increasing $m_H$ as it is correctly 
reproduced in fig. 2. $\tau\rightarrow l_j\mu\mu$
does not receive visible effects by $\delta m/m_{A}$ terms being
dominated by the tree level Higgs exchange.\\
\item
As it is shown in fig. 2, $Br(\tau\rightarrow l_j\gamma)$
is generally larger than $Br(\tau\rightarrow l_j\mu\mu)$;
their ratio is regulated by the following approximate relation:
$$
\frac{Br(\tau\rightarrow l_j\gamma)}{Br(\tau\rightarrow l_j\mu\mu)} \simeq 
\frac{36}{3\!+\!5\delta_{j\mu}}
\frac{Br(\tau\rightarrow l_j\gamma)}{Br(\tau\rightarrow l_j\eta)}
\geq\frac{36}{3\!+\!5\delta_{j\mu}}\,,
$$
where the last relation is valid only out of the cancellation region.\\
Moreover, from the above relation it turns out that:
$$
\frac{Br(\tau\rightarrow l_j\eta)}{Br(\tau\rightarrow l_j\mu\mu)} \simeq 
\frac{36}{3\!+\!5\delta_{j\mu}}\,.
$$
If we relax the condition $\xi_{s,b}= 1$, $Br(\tau\rightarrow l_j\eta)$ can get values few times smaller or bigger than those in fig.2.\\
\item
It is noteworthy that a tree level Higgs exchange predicts that
$Br(\tau\rightarrow l_jee)/Br(\tau\rightarrow l_j\mu\mu)\sim m^2_e/m^2_{\mu}$ 
while, at two loop level, we obtain (out of the cancellation region):
$$
\frac{Br(\tau\rightarrow l_j ee)}{Br(\tau\rightarrow l_j\mu\mu)}\simeq 
\frac{0.4}{3\!+\!5\delta_{j\mu}}
\frac{Br(\tau\rightarrow l_j\gamma)}{Br(\tau\rightarrow l_j\eta)}
\geq\frac{0.4}{3\!+\!5\delta_{j\mu}}\,.
$$
Let us underline that, in the cancellation region, the lower bound of 
$Br(\tau\rightarrow l_jee)$ is given by the monopole contributions.
So, in this region, $Br(\tau\rightarrow l_jee)$ is much less suppressed 
than $Br(\tau\rightarrow l_j\gamma)$.

\end{itemize}

The correlations among the rates of the above processes are an important
signature of the Higgs-mediated LFV and allow us to discriminate between
the gaugino mediated LFV and Higgs-mediated LFV. In fact, in the gaugino mediated case, $Br(\tau\rightarrow l_jl_kl_k)$ get the largest contributions by the dipole amplitudes that are $\tan\beta$ enhanced with respect to all other amplitudes resulting in a precise ratio with $Br(\tau\rightarrow l_j\gamma)$, namely $BR(\tau\rightarrow l_{j}l_{k}l_{k})/BR(\tau\rightarrow l_{j}\gamma)\simeq (\alpha_{el}/3\pi)(\log(m^2_{\tau}/m^2_{l_{k}})-3)$.
Moreover, the gaugino-mediated LFV predicts
$BR(\tau\rightarrow l_{j}ee)/BR(\tau\rightarrow l_{j}\mu\mu)
\simeq(\log(m^2_{\tau}/m^2_{e})-3)/(\log(m^2_{\tau}/m^2_{\mu})-3)\simeq 5$.\\
If some ratios different from the above were discovered,
then this would be clear evidence that some new process is generating the
$\tau\rightarrow l_j$ transition, with Higgs mediation being a leading candidate.
\begin{figure}[h]
\begin{tabular}{cc}
\includegraphics[scale=0.35]{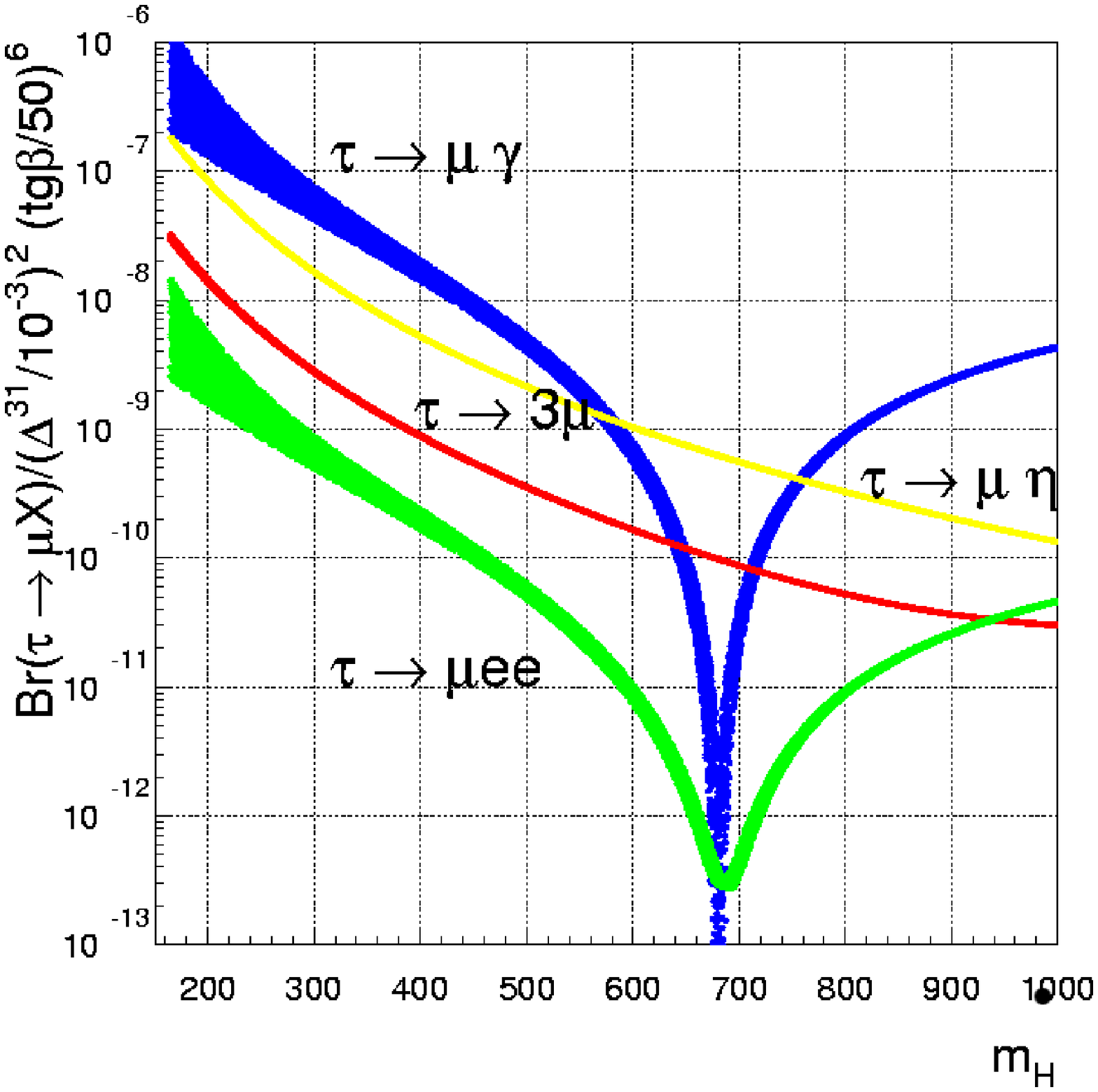} &
\includegraphics[scale=0.35]{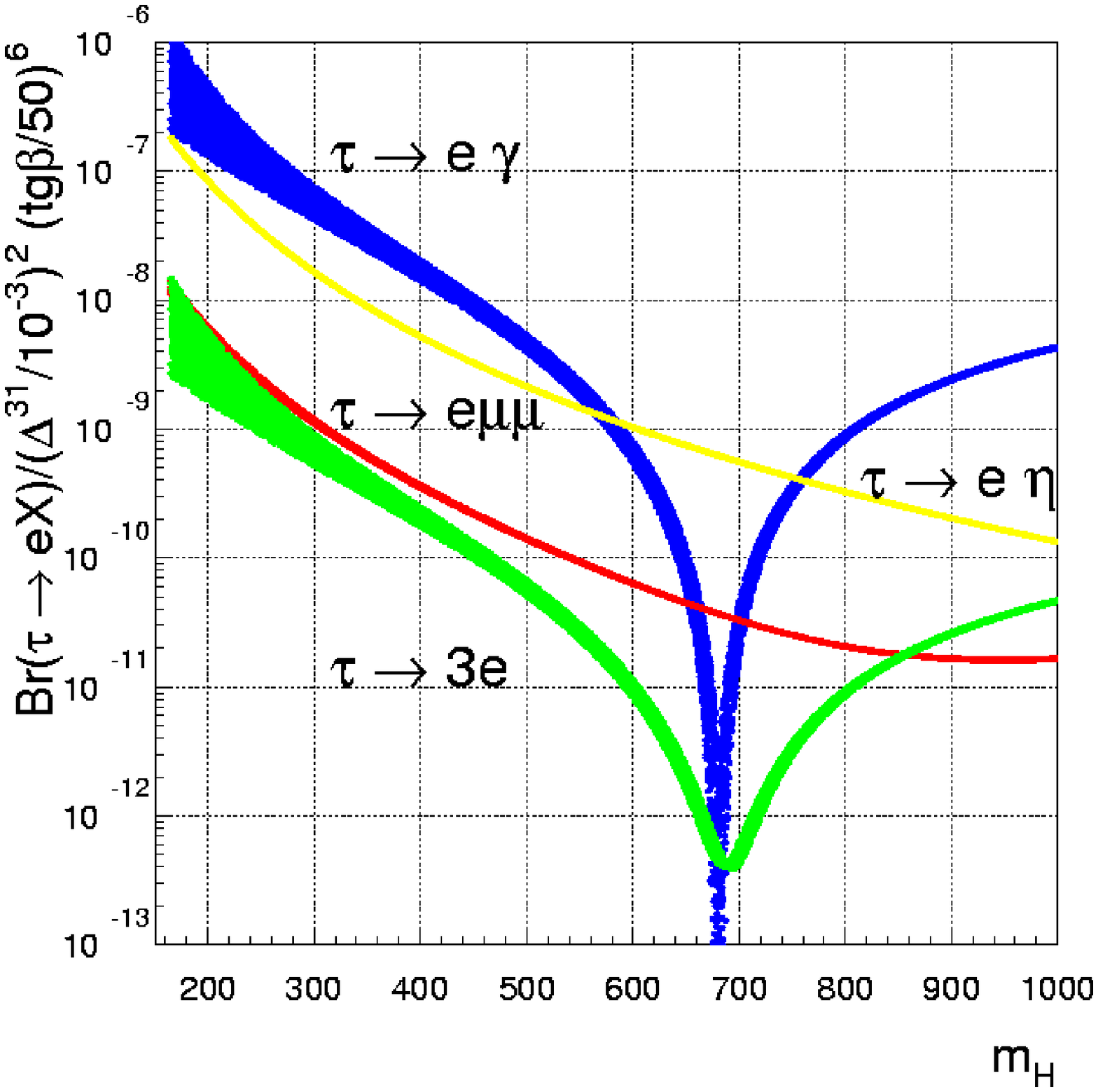}
\end{tabular}
\caption{Branching ratios of various $\tau\rightarrow\mu$ and
$\tau\rightarrow e$ LFV processes vs the Higgs boson
mass $m_H$ in the decoupling limit.
In the figures we assume $X= \gamma, \mu\mu, ee, \eta$. The bands 
correspond to the allowed $\delta m/m_{A}$ values as explained in the text.}
\end{figure}

\section{Conclusions}
In this letter we have studied the allowed rates for Higgs-mediated
LFV decays both in a general two Higgs Model and in Supersymmetry.\\
In particular, we have analyzed the decay modes of the $\tau$ lepton, namely
$\tau\rightarrow l_jl_kl_k$, $\tau\rightarrow l_j\gamma$ and
$\tau\rightarrow l_j\eta$.
Analytical relations and correlations among the rates of the above
processes have been established at the two loop level in the Higgs Boson exchange.\\
The correlations among the processes are a precise signature of the
theory. In this respect experimental improvements in all the decay
channels of the $\tau$ lepton would be very welcome.
We have parametrized the source of LFV in a model independent
way in order to be as general as possible.
We found that $\tau\rightarrow l_j\gamma$ processes are generally the most
sensitive channels to probe Higgs-mediated LFV
specially if the splitting among the neutral Higgs bosons masses is not below $10\%$. 
This condition can be fulfilled if $M_{A(H)}\sim M_W$,
that is, just the situation in which the Higgs LFV effects are more effective.
We have also shown that $\tau\rightarrow l_j\eta$ and $\tau\rightarrow l_jl_kl_k$
are very useful probes of this scenario.
In conclusion, we can say that the Higgs-mediated contributions to
LFV processes can be within the present or upcoming experimental
resolutions and provide an important chance to detect new physics
beyond the Standard Model.\\ \\

\textbf{Acknowledgments:}
I thank A.Brignole, G.Isidori and A.Masiero for useful discussions.


\begin{thebibliography}{999}
\bibitem{fbam}
F.~Borzumati and A.~Masiero,
Phys.\ Rev.\ Lett.\  {\bf 57}, 961 (1986).

\bibitem{Masierorew}
For a recent review see for example,
A.~Masiero, S.~K.~Vempati and O.~Vives,
New J.\ Phys.\  {\bf 6}, 202 (2004)
[arXiv:hep-ph/0407325].

\bibitem{textures}
An incomplete list of references:
F.~Gabbiani, E.~Gabrielli, A.~Masiero and L.~Silvestrini,
Nucl.\ Phys.\ B {\bf 477}, 321 (1996)
[arXiv:hep-ph/9604387];
R.~Barbieri and L.~J.~Hall,
Phys.\ Lett.\ B {\bf 338}, 212 (1994)
[arXiv:hep-ph/9408406];
R.~Barbieri, L.~J.~Hall and A.~Strumia,
Nucl.\ Phys.\ B {\bf 445}, 219 (1995)
[arXiv:hep-ph/9501334];
J.~Hisano, T.~Moroi, K.~Tobe and M.~Yamaguchi,
Phys.\ Rev.\ D {\bf 53} (1996) 2442
[arXiv:hep-ph/9510309];
J.~Hisano, T.~Moroi, K.~Tobe, M.~Yamaguchi and T.~Yanagida,
Phys.\ Lett.\ B {\bf 357}, 579 (1995)
[arXiv:hep-ph/9501407];
J.~Hisano and D.~Nomura,
Phys.\ Rev.\ D {\bf 59}, 116005 (1999)
[arXiv:hep-ph/9810479];
J.~A.~Casas and A.~Ibarra,
Nucl.\ Phys.\ B {\bf 618} (2001) 171
[arXiv:hep-ph/0103065];
S.~Lavignac, I.~Masina and C.~A.~Savoy,
Phys.\ Lett.\ B {\bf 520}, 269 (2001)
[arXiv:hep-ph/0106245];
S.~Lavignac, I.~Masina and C.~A.~Savoy,
Nucl.\ Phys.\ B {\bf 633}, 139 (2002)
[arXiv:hep-ph/0202086];
A.~Masiero, S.~K.~Vempati and O.~Vives,
Nucl.\ Phys.\ B {\bf 649}, 189 (2003)
[arXiv:hep-ph/0209303];
J.~R.~Ellis, M.~E.~Gomez, G.~K.~Leontaris, S.~Lola and D.~V.~Nanopoulos,
Eur.\ Phys.\ J.\ C {\bf 14}, 319 (2000)
[arXiv:hep-ph/9911459];
W.~Buchmuller, D.~Delepine and F.~Vissani,
Phys.\ Lett.\ B {\bf 459}, 171 (1999)
[arXiv:hep-ph/9904219];
W.~Buchmuller, D.~Delepine, L.T.~Handoko 
Nucl.\ Phys.\ B {\bf 576},445 (2000)
[arXiv:hep-ph/9912317];
Y.~Kuno, Y.~Okada 
Rev.\ Mod.\ Phys.\ {\bf 73},151 (2001)
[arXiv:hep-ph/9909265].

\bibitem{mfvl}
V.~Cirigliano, B.~Grinstein, G.~Isidori and M.~B.~Wise,
arXiv:hep-ph/0507001.

\bibitem{Hrs}
L.~J.~Hall, R.~Rattazzi and U.~Sarid,
Phys.\ Rev.\ D {\bf 50}, 7048 (1994)
[arXiv:hep-ph/9306309];
T.~Blazek, S.~Raby and S.~Pokorski,
Phys.\ Rev.\ D {\bf 52}, 4151 (1995)
[arXiv:hep-ph/9504364];
M.~Carena, D.~Garcia, U.~Nierste and C.~E.~M.~Wagner,
Phys.\ Lett.\ B {\bf 499}, 141 (2001)
[arXiv:hep-ph/0010003].

\bibitem{bkq}
K.~S.~Babu and C.~F.~Kolda,
Phys.\ Rev.\ Lett.\  {\bf 84}, 228 (2000)
[arXiv:hep-ph/9909476].

\bibitem{isiburapila}
G.~D'Ambrosio, G.~F.~Giudice, G.~Isidori and A.~Strumia,
Nucl.\ Phys.\ B {\bf 645}, 155 (2002)
[arXiv:hep-ph/0207036];
G.~Isidori and A.~Retico,
JHEP {\bf 0111}, 001 (2001)
[arXiv:hep-ph/0110121];
G.~Isidori and A.~Retico,
JHEP {\bf 0209}, 063 (2002)
[arXiv:hep-ph/0208159];
A.~J.~Buras, P.~H.~Chankowski, J.~Rosiek and L.~Slawianowska,
Nucl.\ Phys.\ B {\bf 659}, 3 (2003)
[arXiv:hep-ph/0210145];
A.~Dedes,
Mod.\ Phys.\ Lett.\ A {\bf 18}, 2627 (2003)
[arXiv:hep-ph/0309233];
A.~Dedes and A.~Pilaftsis,
Phys.\ Rev.\ D {\bf 67}, 015012 (2003)
[arXiv:hep-ph/0209306].

\bibitem{bkl}
K.~S.~Babu and C.~Kolda,
Phys.\ Rev.\ Lett.\  {\bf 89}, 241802 (2002)
[arXiv:hep-ph/0206310].

\bibitem{chang1}
D.~Chang, W.~S.~Hou and W.~Y.~Keung,
Phys.\ Rev.\ D {\bf 48}, 217 (1993)
[arXiv:hep-ph/9302267];

\bibitem{chang}
M. Sher and Y. Yuan, Phys.\ Rev.\ D {\bf 44}, 1461 (1991);
R.~A.~Diaz, R.~Martinez and J.~A.~Rodriguez,
Phys.\ Rev.\ D {\bf 64} (2001) 033004
[arXiv:hep-ph/0103050];
R.~Diaz, R.~Martinez and J.~A.~Rodriguez,
Phys.\ Rev.\ D {\bf 63}, 095007 (2001)
[arXiv:hep-ph/0010149];
Y.~F.~Zhou,
J.\ Phys.\ G {\bf 30}, 783 (2004)
[arXiv:hep-ph/0307240];
S.~Kanemura, T.~Ota and K.~Tsumura,
arXiv:hep-ph/0505191;
S.~N.~Gninenko, M.~M.~Kirsanov, N.~V.~Krasnikov and V.~A.~Matveev,
Mod.\ Phys.\ Lett.\ A {\bf 17}, 1407 (2002)
[arXiv:hep-ph/0106302];
M.~Sher and I.~Turan,
Phys.\ Rev.\ D {\bf 69}, 017302 (2004)
[arXiv:hep-ph/0309183].


\bibitem{sher}
M.~Sher,
Phys.\ Rev.\ D {\bf 66}, 057301 (2002)
[arXiv:hep-ph/0207136];
A.~Dedes, J.~R.~Ellis and M.~Raidal,
Phys.\ Lett.\ B {\bf 549}, 159 (2002)
[arXiv:hep-ph/0209207];
A.~Brignole and A.~Rossi,
Phys.\ Lett.\ B {\bf 566}, 217 (2003)
[arXiv:hep-ph/0304081];
A.~Brignole and A.~Rossi,
Nucl.\ Phys.\ B {\bf 701}, 3 (2004)
arXiv:hep-ph/0404211; 
E.~Arganda, A.~M.~Curiel, M.~J.~Herrero and D.~Temes,
Phys.\ Rev.\ D {\bf 71}, 035011 (2005)
[arXiv:hep-ph/0407302];
S.~Kanemura, K.~Matsuda, T.~Ota, T.~Shindou, E.~Takasugi and K.~Tsumura,
Phys.\ Lett.\ B {\bf 599}, 83 (2004)
[arXiv:hep-ph/0406316];
S.~Kanemura, Y.~Kuno, M.~Kuze and T.~Ota,
Phys.\ Lett.\ B {\bf 607}, 165 (2005)
[arXiv:hep-ph/0410044];
R.~Kitano, M.~Koike, S.~Komine and Y.~Okada,
Phys.\ Lett.\ B {\bf 575}, 300 (2003)
[arXiv:hep-ph/0308021].

\bibitem{hlfvemu}
P. Paradisi, arXiv:hep-ph/0601100.

\bibitem{chengsher}
T.P. Cheng and M. Sher, Phys.\ Rev.\ D {\bf 35}, 3484 (1987).

\bibitem{fedelman}
T. Fedelman,
Int.\ J. Mod.\ Phys.\ A\ {\bf 15} (2000) 159, [arXiv:hep-ph/9907491].

\bibitem{haber}
H. E. Haber and R. Hempfling,
Phys.\ Rev.\ D {\bf 48}, 9, 4280 (1993).

\bibitem{1mio}
P.~Paradisi,
JHEP {\bf 0510}, 006 (2005)
[arXiv:hep-ph/0505046];
M. Ciuchini, A. Masiero, P. Paradisi, L. Silvestrini, S. K. Vempati and O. Vives,
\textit{to appear};
I.~Masina and C.~A.~Savoy,
Nucl.\ Phys.\ B {\bf 661}, 365 (2003)
[arXiv:hep-ph/0211283].


\end{thebibliography}
\end{document}